# Electrostatic Deflection of a Molecular Beam of Massive Neutral Particles: Fully Field-Oriented Polar Molecules within Superfluid Nanodroplets

*Daniel J. Merthe and Vitaly V. Kresin\**

Department of Physics and Astronomy, University of Southern California,
Los Angeles, CA 90089-0484, USA

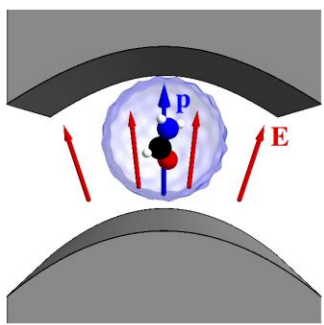

**Abstract**

Electric deflection measurements on liquid helium nanodroplets doped with individual polar molecules demonstrate that the cold superfluid matrix enables full orientation of the molecular dipole along the external field. This translates into a deflection force that is increased enormously by comparison with typical deflection experiments, and it becomes possible to measurably deflect neutral doped droplets with masses of tens to hundreds of thousands of Daltons. By using continuous fluxes of fully oriented polar molecules and measuring the deflection of the doped nanodroplet beam, this approach makes it possible to directly determine the dipole moments of internally cryogenically cold molecules. The technique is broadly and generally applicable, including to complex and biological molecules.

Inspired by the progress and insights that arose out of cooling and trapping experiments on atoms, there has been a burst of interest in slowing, cooling and manipulating molecules.[1-3] Polar molecules have attracted particular attention because their electric dipole moment provides a "handle" with which the they can be slowed, focused, oriented, and otherwise manipulated by external fields and laser pulses. Dipole moments also give rise to strong interparticle interactions which can facilitate novel reaction pathways and quantum entanglement effects. In fact, the use of electric fields to orient and steer polar molecules has a long pedigree: the first electric deflection experiments on "molecular rays" were performed in the late 1920s to early 1930s in O. Stern's laboratory.[4,5]

Unfortunately, it is not easy to impart a strong orientation even to highly polar molecules in the gas phase. The problem derives from the molecules' rotational motion which washes out the effect of the external field's torque. Indeed, it is well known from Langevin-Debye theory[6] that polarization $\langle p_z \rangle$ (i.e., the average projection of the dipole $p$ on the direction of the electric field $E_z \hat{z}$) is characterized by the parameter $\alpha \equiv pE_z / k_B T$, where $T$ is the rotational temperature of the ensemble. In the linear regime when $\alpha \ll 1$, $\langle p_z \rangle$ is only a fraction $pE_z / (3k_B T)$ of the molecules' total dipole moment. For example, for water molecules ($p$=1.85 D), at room temperature and within a strong laboratory field of 100 kV/cm, this fraction is a paltry $5\times10^{-3}$.

With the use of supersonic molecular jets, especially pulsed ones, rotational temperatures below 1 K have been demonstrated[7,8] but temperatures of several K are much more representative. Pendular states of molecules in such cold expansions have been investigated extensively,[9] but impractically large *dc* electric fields would be required to obtain nearly-full orientation in most cases.[10] Note also that vibrational degrees of freedom are not cooled as efficiently in supersonic expansions.[11] While much stronger *ac* electric fields exist within pulsed laser beams and have been applied to molecular orientation,[12] their spatial and temporal extent is small, and they carry the risk of resonant or nonresonant internal excitation and dissociation of the molecule. Electrostatic multipole focusing[13,14] can impose a degree of orientation on molecular beams, but these bulky (sometimes meters-long) rod assemblies are limited in the selection of usable quantum states and molecules. Similarly, electrostatically guided[15,16] or hydrodynamically enhanced[17] beams of small molecules from a buffer-gas cooled reservoir are fruitful sub-Kelvin options, but again here one has to contend with isolating quantum states susceptible to guidance and with efficiently entrapping molecules in the flow. Critically, the aforementioned methods are challenged when it comes to larger and heavier polar molecules, not to mention organic or biological ones.

Thus, it is valuable to develop techniques capable of all of the following: (i) deliver robust fluxes of highly oriented cold molecules, (ii) be broadly applicable to many types of molecules, (iii) not require specialized intense lasers, (iv) exhibit the orientation by an unambiguous direct measurement, (v) utilize this orientation for the exploration of physical properties and phenomena.

We present here an approach that meets the above targets. The core idea is to embed polar molecules within very cold superfluid helium nanodroplets in a collimated supersonic beam. The doped nanodroplets then pass through an inhomogeneous electrostatic field. The measurements demonstrate that the molecules become fully field-oriented within the droplets, resulting in an exceptionally strong deflecting force.

"Helium nanodroplet isolation" ("HENDI")[18-21] is a versatile tool for spectroscopy and mass spectrometry. Molecules and atoms are picked up, entrapped, cooled, and transported by a beam of $^4$He$_N$ nanodroplets generated by supersonic expansion of pure helium gas through a cryogenic nozzle. The average droplet size $<N>$ can be adjusted from $\sim 10^2$ to $>10^6$ by tuning the expansion conditions. Upon leaving the source, the nanodroplets promptly undergo evaporative cooling to a final internal temperature of only 0.37 K and turn superfluid.

As the droplets collide with background atoms and molecules while passing through a cell (or a series of cells) the latter are readily picked up, cooled by prompt heat transfer to the helium matrix (followed by the evaporation of additional helium atoms which brings the droplet back to its original temperature), and carried along by the droplet beam. The probability of a sticking encounter within the pick-up cell is given by Poisson statistics, and cell pressures in the $10^{-6}$-$10^{-4}$ Torr range are sufficient to result in the majority of nanodroplets in the beam becoming doped with one or more impurity. The resulting flux of cryogenically cooled molecules is typical for supersonic molecular beams and is within the range achieved by buffer-gas cooling/extraction methods: characteristic droplet fluxes[19] are $\sim 10^{10}$ s$^{-1}$ corresponding to $\sim 10^8$ droplets (and their cold molecules) per cm$^3$.

It must be re-emphasized that nanodroplet embedding is applicable to a very wide range of molecules: all that is required is to generate a low vapor pressure sufficient for pick-up. It is also possible to coembed two or more units and thus produce unusual or metastable polar complexes that would be unobtainable by other means. In addition, the helium environment efficiently cools not only the rotational, but also the vibrational degrees of freedom of the embedded impurity. With buffer gas methods this can be a challenge.

With the exception of some alkali and alkaline-earth atoms and clusters,[21,22] the picked-up dopants sink into the droplet. Rotational spectroscopy (see, e.g., the aforementioned reviews in refs. 18, 19, and 21) has confirmed that dopants rotate freely in the superfluid nanodroplet environment. Therefore, they also can be easily oriented by an external field. This has already been made use of by landmark experiments on the spectroscopy of embedded species in pendular states.[23] The new step taken in our work is that the droplet beam's full trajectory is measurably deflected by the application of an inhomogeneous field. We demonstrate that this effect can be employed to infer quantitative information about the system, and, importantly, that it can be used for complex (including biological) as well as simple molecules without the prerequisite of potentially difficult spectroscopic analysis.

As mentioned earlier, deflection of molecular beams by static field gradients is a longstanding experimental tool. More recently, it has been applied to beams of nanoclusters (see, e.g., the reviews in refs. 24-27). However, employing it with helium droplets may appear questionable. After all, if the deflection of a beam of individual polar molecules or clusters in a typical experiment is at most a few millimeters, or more commonly a fraction of a millimeter, then what could possibly be detected after adding to them many tens of thousands of Daltons of barely polarizable helium liquid?

The answer, as already emphasized, lies in the molecular orientation enabled by nanodroplet embedding. The resulting increase in the deflecting force is so great that it easily compensates for the enlarged mass.

Indeed, each nanodroplet acts as a thermal bath for its molecule and we can use the full Langevin function to find the dipole orientation: $\langle p_z \rangle = p[\coth\alpha - 1/\alpha]$, where $T$=0.37 K and $\alpha$ was defined above.[28] Typical electric deflection measurements above take place in the $\alpha \ll 1$ regime, and then $\alpha/3$ becomes the aforementioned suppression factor. However, now we are in the opposite limit, the Langevin function is almost fully saturated,[30] and this enables an orders-of-magnitude increase in the field-aligned dipole component and a corresponding increase in the deflecting force. (For most deflector designs the field gradient, which provides the deflecting force, is collinear with the field itself.)

For example, the dimethyl sulfoxide (DMSO) molecule used as a reference dopant in the measurements described below has a dipole moment of 4 D units and acquires an alignment $<p_z>/p \approx 0.96$ at $E$=100 kV/cm, using the classical Langevin expression. A quantum-mechanical calculation employing the rotational constants of the free molecule reduces this alignment to 0.87, still a very high value. For comparison, the current record for laser alignment is 60% using the ground-state OCS molecule in a 0.5-ns-duration high-intensity laser pulse.[32]

Note that the discussion above is, strictly speaking, applicable for molecules with sufficiently small rotational constants: the classical Langevin formula can be employed if a considerable number of rotational levels are occupied at 0.37 K. In the opposite case (for example, the rotational constants of $NH_3$ are 9 K and 14 K) one needs to consider the individual rotational Stark levels of the molecules, but the general conclusion about very strong orientation is sustained when quantum corrections are included.[33]

Electrostatic beam deflection measurements on doped nanodroplets were performed in an apparatus described in the Supporting Information. A collimated $^4He_N$ beam with a narrow velocity distribution picked up DMSO, formamide, or histidine molecules under conditions optimized to an average of one dopant molecule per droplet. It then passed through a gap between two electrodes where a strong inhomogeneous electric field oriented the dopant and exerted a sideways force. The resulting deflection was determined by measuring the beam profile at the entrance to a mass spectrometer tuned to the most intense characteristic fragment peak of the molecular impurity.

The electric field produced by the plates and its gradient are both proportional to the applied voltage $V$. The deflection angle of a nanodroplet during its flight to the detector is proportional to the ratio of the sideways impulse it receives from the field, $F_z \Delta t \propto p_z (\partial E / \partial z) / \upsilon$, to its original forward momentum, $m\upsilon$. Therefore its deflection $d$ in the detector plane is $d = C p_z V / (m \upsilon^2)$, where $C$ is an overall geometrical coefficient.

The formula points out a hurdle that arises when the deflection method is applied to doped nanodroplets. Whereas in conventional experiments the mass of the deflected particle is exactly specified, helium beams contain a spread of nanodroplet sizes $N$. As a result, the polar molecules $M$ arrive at the detector riding within droplets of various sizes: $M@He_N$. In subcritical expansion conditions, such as those employed here, the droplet size distribution $P(N)$ is log-normal with the average size and the standard deviation related as $\Delta N \approx 0.65 <N>$ (as shown by the data in ref. 34). Therefore the deflection profile is the result of convolving $d$ with the broad droplet mass distribution. Notwithstanding, it is possible to obtain quantitative data about the magnitude of embedded molecular dipoles. The procedure is to measure the deflection profile of a beam doped with a molecule whose dipole moment is known, and then, keeping the source conditions and therefore the droplet ensemble unchanged, the deflection of a beam with the molecule of interest. The difference in deflections is then directly related to the ratio of the two dipole moments.

In practice, we implemented this procedure via a Monte Carlo simulation consisting of the following steps. A nanodroplet is picked from a log-normal size distribution, and is assigned a velocity $v$ from the starting narrow longitudinal and transverse distributions. The energy of the collision with a dopant molecule ($\frac{1}{2}m\upsilon^2$) plus its thermal translational and rotational motion ($3k_B T_c$, where $T_c$ is the vapor temperature in the pick-up cell) is dissipated by the evaporation of He atoms from the droplets (0.5-0.6 meV energy release per He atom,[18,35] typically a few hundred atoms are lost). As regards the internal molecular energies, their characteristic vibrations in the compounds studied in this Letter lie $\gtrsim$100 meV (1000 K) in energy, and therefore are not significantly excited at $T_c$.[36]

The doped nanodroplet, cooled by prompt evaporation back to 0.37 K, then enters the electric field region where the orientation $<p_z>/p$ is sampled from the in-field Boltzmann distribution. The field and the embedded dipole also weakly polarize the helium matrix; this is not a large contribution to the deflection pattern but is accounted for in the simulation (see Supporting Information). The electric plates' calculated field gradient[37] is used to compute the deflection force on the droplet during its passage through the field. The droplet is then allowed to drift through the free flight path until it arrives at the detector entrance. This procedure was repeated $10^6$ times until a simulated beam profile was generated. It was then convolved with the transmission function of the scanning slit for comparison with the experimental data, the fitting parameters being the average droplet size <N> and the molecular dipole moment $p$.

Figure 1 shows the experimental results for three molecules chosen to illustrate the method. The dots represent experimental data, and the lines are simulation fits according to the procedure described above. The area under the measured profiles decreases at higher deflection voltages, possibly due to a larger-than-expected fraction of the smallest droplets in $P(N)$ which are swept outside the detection region. The amplitudes of the fitted curves are adjusted to the experimental values. As the reference molecule, Fig. 1(a), we used DMSO, $(CH_3)_2SO$, with a 4.0 D dipole moment,[38] and the average nanodroplet size deduced from this profile was <N>≈25,000, in good agreement with the literature value[18] for the source parameters used in the experiment. This supports the consistency of the approach.

Using the same source conditions in a different run, we measured the deflections of a droplet beam doped with formamide ($CH_3NO$), a tabulated dipole moment of 3.7 D.[38] The fit to the experimental data in Fig. 1(b) yielded 3.2±0.3 D. This present accuracy of ≈15% will be improved with the planned upgrades to higher deflecting voltages and ion counting sensitivity.

The third selected molecule, histidine, is an aromatic amino acid involved in important photochemical, proton transfer, and metal binding processes.[39-43] The mass spectrometry of its complexes in He nanodroplets has been explored experimentally.[44,45] Not long ago, its conformational topology was explored at the *ab initio* level, including a calculation of the electric dipole moments.[46] The ground-state conformer was predicted to be dominant in the gas phase (and therefore the one that should be captured and interrogated by HENDI), relatively rigid and compact, and to have a dipole moment of 4.9 D. The deflection profile shown in Fig. 1(c) yields a value of 3.0±0.6 D, which implies, even considering the current experimental precision, that the theoretical calculation has overestimated the dipole moment. The amino acid dipole moments are sensitive to the behavior of side chain orientations (rotamers), and the present data suggests that there is need for additional theoretical analysis.

The key feature of the experiment is seen in the insets in Fig. 1. They show that the deflections $d$ are proportional to the applied voltage $V$, i.e., to the electric field $E_z$. This is fundamentally different from what has been seen previously for the electric deflections of nanoclusters possessing pure polarizability or linear (induced) susceptibility.[24-26,47] In those cases $p_z \propto E_z$, as discussed above, and therefore it has always been observed that $d \propto V^2$. The fact that in the present case $d \propto V$ represents unequivocal proof that saturated orientation of the molecular dipoles has been achieved.

In summary, we have demonstrated an efficient and general approach to the preparation and study of a continuous flux of fully oriented polar molecules. Individual molecules are picked up by passing liquid helium droplets and find themselves carried along by an inert, superfluid, and very cold nanoscale matrix where they become thermalized to below 0.4 K and are freely able to rotate and reorient. As the doped nanodroplet beam is passed through an electric field region, the embedded molecules become aligned along the field and in this way experience a vastly enhanced steady deflecting force. As a result, we are able to induce and resolve the electrostatic deflection of neutral systems with unprecedentedly high masses of tens to hundreds of thousands of Daltons.

The technique has a number of novel and useful features and applications. Importantly, it is generally applicable to complex as well as simple molecules. Their electric dipole moments can be determined by a direct measurement, not requiring acquisition and analysis of an optical spectrum.

It would be straightforward to extend the approach to magnetic deflection of nanodroplets doped with superparamagnetic molecules and clusters. For example, a $10^4$-atom helium droplet containing a single Bohr magneton oriented in a 350 T/m field gradient[48] would be deflected by 2-3 mm. Interesting dynamics also may appear if the embedded oriented molecules are additionally subjected to intense polarized aligning laser pulses.[49] Furthermore, sequential doping can produce unusual and peculiar polar metastable configurations and assemblies (see the review in ref. 23) which can be subjected to deflection measurements. In this way it may become possible to make use of the superfluid droplet environment to explore prototypes of novel dipole-dipole interaction physics.

The presently demonstrated ability to embed and field-orient molecules of biochemical interest not only makes it possible to study their configurations via deflectometry (cf. ref. 24) but also opens the door to keeping them aligned for gas-phase crystallographic structure determination.[50,51] Finally, as described in more detail in the Supporting Information, the appearance of large deflections creates a new opportunity to separate neutral $^4$He$_N$ nanodroplets by size, e.g. for the study of size effects in dopant spectroscopy.

**Supporting Information.** Description of the experimental apparatus and methods. Calculation of the helium nanodroplet polarization. Simulations and discussion of nanodroplet size selection using the deflection technique.

**Acknowledgment.** This work was supported by the U.S. National Science Foundation under grant CHE-1213410.

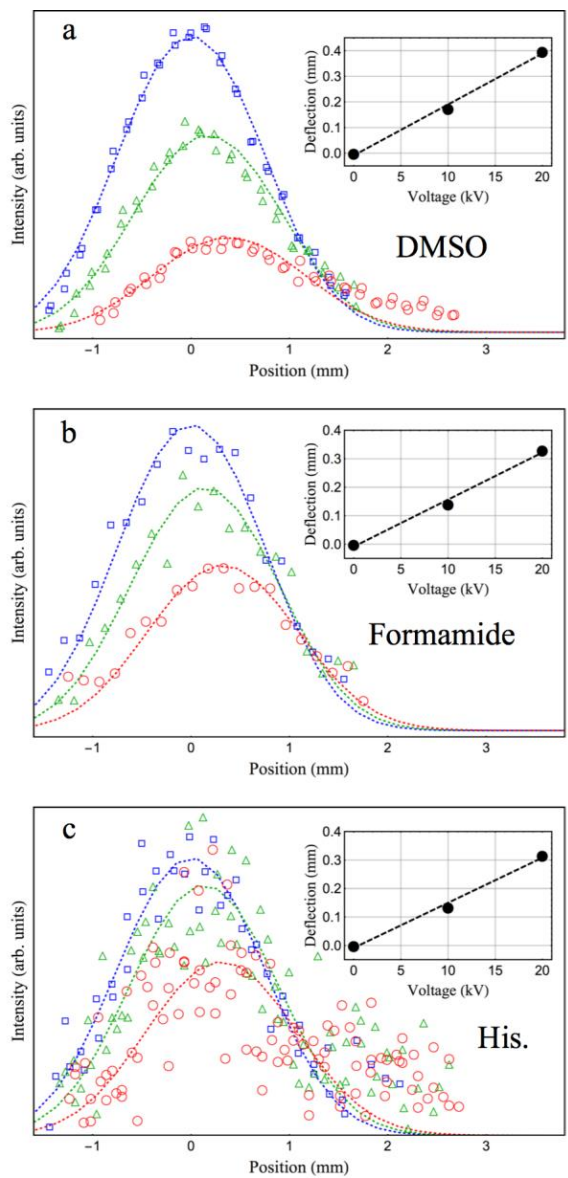

**FIG. 1.** Deflection beam profiles of helium nanodroplets doped with a molecule of (a) DMSO, (b) formamide, (c) histidine. The dots are the experimental data points and the lines are fitted simulated profiles, as described in the text, for deflection voltages of 0 kV (blue), 10 kV (green), and 20 kV (red). For the latter voltage the electric field and gradient experienced by the doped helium droplets are ≈80 kV/cm and ≈270 kV/cm$^2$, respectively. Insets show the profile centroid as a function of applied voltage. The linear dependence attests to the fact that the embedded polar molecules are effectively fully oriented by the applied field.

# Electrostatic Deflection of a Molecular Beam of Massive Neutral Particles: Fully Field-Oriented Polar Molecules within Superfluid Nanodroplets

Daniel J. Merthe and Vitaly V. Kresin

*Department of Physics and Astronomy, University of Southern California,
Los Angeles, CA 90089-0484, USA*

## I. Experiment

The helium nanodroplet apparatus produced a supersonic $^4$He$_n$ beam by expansion of high purity helium gas through a 5 μm nozzle at 40 bar stagnation pressure and 13 K temperature. After passing through a skimmer and a mechanical wheel chopper, the droplets picked up the dopant molecules within a copper cell. The data described below were acquired with formamide or DMSO liquids whose vapor was fed from an outside glass container through a capillary (with its flow kept constant using a regulated combination of thermally insulating tape, fine needle valves, and differential pumping with a mechanical vane pump) and with histidine which was loaded into the cell as a powder and heated with the use of a temperature controller. In all cases, the dopant's partial pressure in the pickup cell was adjusted to maximize the signal of characteristic single-molecule ionization products, as identified by the mass spectrometry tables[S1]. For Poissonian pickup statistics this point corresponds to an average of one dopant molecule per droplet. A small fraction of larger dopant clusters may also produce single-molecule ions and their fragments in the mass spectrum, however, based on the good fits of mass spectral line intensities to Poisson distributions of dopant populations in many studies (e.g., refs. S2 and S3) this contribution can be neglected at the present level of precision. The stability of the doped beam was verified by monitoring the intensities of the peaks in the mass spectrum.

In the following chamber, the beam was collimated by a 0.25 mm × 1.25 mm slit and entered the 2.5 mm wide gap between two 15 cm long metal plates shaped to create a "two-wire" inhomogeneous electric field[S4,S5]. The field oriented the polar dopant molecule and its gradient exerted a deflecting force on the oriented dipole. By applying voltage of up to 20 kV between the plates, electric fields up to 80 kV/cm and field gradients up to 250 kV/cm$^2$ were created.

The droplet beam was detected by a Balzers QMG-511 crossed-beam quadrupole mass analyzer with an electron impact ionization source set to an energy of 70 eV. For each beam deflection measurement the analyzer was tuned to the most intense characteristic fragment peak of the molecular impurity[S1]. Selecting the monomer peak in the mass spectrum mitigates influences from larger dopants agglomerates on the measured deflections. The output of the analyzer was fed into a lock-in amplifier together with the chopper synchronization pulses, filtering out the background and extracting the signal carried by the helium beam.

In addition, the lock-in's phase delay between the chopper pulse and the analyzer output was used to determine the velocity of the supersonic droplet beam. It was found to be $v_0$=400 m/s. The distribution of forward velocities is narrow; according to ref. S6 its width can be taken as $\Delta v \approx 0.03\ v_0$. To account for beam divergence one also can assign it a transverse Gaussian velocity distribution with a mean of zero and a standard deviation of $4\times10^{-4}\cdot v_0$, as parametrized from the zero-field beam profiles. This is consistent with what would be expected from a conical projection of the molecular beam from the skimmer through the collimator.

The lock-in was read by a LabVIEW computer program, which also moved a 0.25 mm wide slit in front of the ionizer entrance, a distance $L$=140 cm past the middle of the deflection field plates. This slit, scanned across the beam by a stepper motor in steps of 0.15 mm, sampled the beam intensity at 20-25 slit positions in a sequence that was randomized for each pass across the beam. Each voltage-on and voltage-off profile typically combined approximately 100 passes for a total acquisition time of 2-3 hours per profile.

## II. Helium droplet polarization

To estimate the polarization induced in the helium nanodroplet by the external field and by the embedded polar molecule, we approximated it as a spherical dielectric shell with dielectric constant $\varepsilon = 1.057$[S7], an outer radius of $b = 0.2\times N^{1/3}$ nm[S8] and an inner radius of $a = 1$ nm, with the inner surface enclosing the dopant molecule.

The total dipole moment of the spherical shell will depend only on the dipole field of the embedded molecule and not on any other multipoles of its charge distribution. Consequently, the problem can modeled as that of an electrostatic shell with a point dipole $p$ at the center and a uniform electric field $E\hat{z}$ applied externally. We assume that the embedded dipole is oriented in the direction of the external field.

By the standard method of polynomial expansion and boundary condition matching, one can calculate the electrostatic potential, the shell polarization $P\hat{z}$ and, by integrating the latter, the total induced dipole moment of the nanodroplet (in Gaussian units):

$$p_z^{He} = \left(1-\frac{a^3}{b^3}\right)\frac{b^3(2\varepsilon+1)(\varepsilon-1)E-2(\varepsilon-1)^2 p}{(2\varepsilon+1)(\varepsilon+2)-2(a/b)^3(\varepsilon-1)^2} \qquad (S.1)$$

For an electric field $E$ of 80 kV/cm applied to a nanodroplet containing 25,000 helium atoms and a $p$=4 D impurity, the calculated induced dipole moment is 1 D. Since this effect is an added shift to all doped droplet deflections, its influence on the comparison of embedded molecular dipoles is relatively minor.

## III. Nanodroplet size filtering

The fact that the doped droplets are composed of an ensemble of sizes is both a complication and an opportunity. The complication, as described in the Letter, arises from the need to perform a convolution of the deflection over the droplet size distribution $P(N)$. On the other hand, if the droplets are doped with a molecule possessing a sufficiently high dipole moment, a deconvolution of the neutral droplet size distribution can be performed by the deflection process itself. Since light droplets deflect more than heavy ones, spatial filtering of the deflected beam imparts a size bias onto the transmitted doped droplet population.

This is useful, because so far size selection has been practical only for charged nanodroplets, achieved via electron or ion doping (e.g., see refs. S9 and S10). (Neutral droplet diffraction (e.g., see ref. S11) can be used only for the smallest droplets, while droplet deflection through the impact of atoms from a secondary beam (e.g., see ref. S12) is a very low-yield process.) Therefore the proposed ability to experimentally emphasize specific size segments of the deflected beam ensemble is an attractive possibility. For example, it can facilitate studies of dopants' laser or ionization spectroscopy as a function of their host droplet sizes.

To illustrate the discussion, Fig. S.1 shows the result of a deflection simulation, as described in the main text, of a beam of nanodroplets doped with CsI molecules ($p = 12$ D). The top panel shows the composition of the deflected beam, and the bottom panel shows its relative size enrichment $R$, that is, the ratio of the proportion of droplets of a certain size $N$ found at various positions in the detector plane to the original proportion of that size in the nozzle beam:

$$R(x; N, \Delta N) = \frac{\left. \int_{N}^{N+\Delta N} I(x;N)dN \middle/ \int_{0}^{\infty} I(x;N)dN \right.}{\int_{N}^{N+\Delta N} P(N)dN} . \qquad (S.2)$$

Here $I(x;N)$ is the intensity of doped nanodroplets of size $N$ at the position $x$ in the detector plane. The figure shows that it is realistic to locate positions in the deflected beam where specific constituent sizes become more prominent or even dominant.

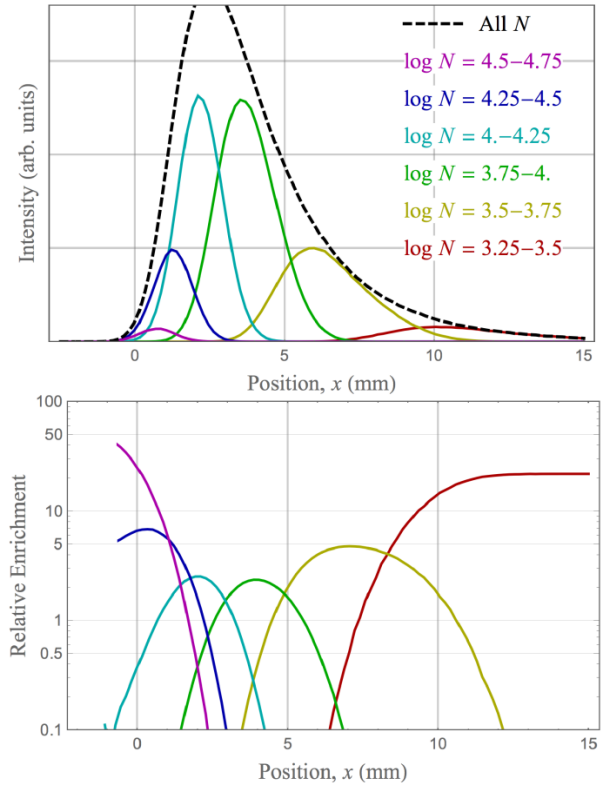

**FIG. S.1.** Simulated deflection of a beam of He nanodroplets with a log-normal size distribution with the average size $\langle N \rangle = 10^4$ and width $\Delta N = 6500$, doped with CsI. The undeflected beam profile has a width of approximately 1 mm. A 100 kV/cm field is applied to the deflecting plates. Lines of different colors correspond to different segments of the droplet size distribution arriving at the detector. Panel (a) shows the total beam profile and its underlying size composition, and panel (b) shows the relative enrichment of different sizes as a function of deflection position in the detector plane, Eq. (S.2).